\documentclass[english,preprint]{aastex}
\usepackage[T1]{fontenc}
\makeatletter
\usepackage{babel}
\makeatother
\begin{document}

\title{Probing the magnetic field with molecular ion spectra - II}

\author{Martin Houde\altaffilmark{1,2},\email{houde@ulu.submm.caltech.edu}
Ruisheng Peng\altaffilmark{1}, Thomas G. Phillips \altaffilmark{3},
Pierre Bastien\altaffilmark{2}, \and Hiroshige Yoshida \altaffilmark{1}}

\altaffiltext{1}{Caltech Submillimeter Observatory, 111 Nowelo Street, Hilo, HI 96720}

\altaffiltext{2}{Département de Physique, Université de Montréal, Montréal, Québec H3C 3J7, Canada}

\altaffiltext{3}{California Institute of Technology, Pasadena, CA 91125}

\begin{abstract}
We present further observational evidence in support of our earlier
proposal \citep{Houde 2000} for detecting the presence of the magnetic
field in molecular clouds by comparing spectra of molecular ions with
those of neutral molecules. The ion lines tend to be narrower and
do not show the wings due to flows, when the magnetic field is sufficiently
strong. We obtained spectra for the optically thin lines of the H$^{13}$CN
and H$^{13}$CO$^{+}$ species in a sample of ten molecular clouds
and found the results to be in agreement with our previous observations
of the main isotopic species, HCN and HCO$^{+}$, made in OMC1, OMC2,
OMC3 and DR21OH, thus eliminating the possibility of optical depth
effects playing a role in the ion line narrowing. HCS$^{+}$ was also
detected in four of these star forming regions. We also discuss previously
published results by \citet{Benson 1998} of N$_{2}$H$^{+}$ detections
in a large sample of dark clouds. We show that the similarity in line
widths between ion and neutral species in their sample is consistent
with the relatively small amount of turbulence and other flows observed
in these clouds.
\end{abstract}

\keywords{ISM: cloud --- ISM: magnetic field --- ISM: molecules}

\section{Introduction.\label{sec:intro}}

In a previous paper \citep[hereafter Paper I]{Houde 2000}, we argued
that the presence of a magnetic field in a weakly ionized plasma can
be detected through comparisons between ion and neutral molecular
line profiles. It was shown that, for regions inhabited with a sufficiently
strong magnetic field, we expect ions to exhibit narrower line profiles
and a significant suppression of high velocity wings in cases where
there is non-alignment between the mean field and the neutral flow(s).
A further condition is that the flow(s) involved (which could also
exist in the form of vorticity, i.e. turbulence) must not have a local
zero mean velocity. For a Maxwellian gas velocity distribution there
will be no effect.

As supporting evidence for our assertions, we presented observations
of HCN, HCO$^{+}$ and N$_{2}$H$^{+}$ for the OMC1, OMC2, OMC3 and
DR21OH molecular clouds obtained at Caltech Submillimeter Observatory
(CSO). In each instance, we observed the effect described above. One
could, however, argue that if for some reason the lines of the neutral
species were to be significantly more opaque than their ionic counterparts,
the differences in the line profiles between the neutrals and ions
could simply be the result of greater saturation of the neutral lines
rather than the presence of the magnetic field. It would then appear
necessary to verify the effect on less abundant species which are
more likely to exhibit optically thin lines. This paper addresses
this question by presenting detections of H$^{13}$CN and H$^{13}$CO$^{+}$
in a group of ten molecular clouds, including the four studied in
Paper I. Furthermore, HCS$^{+}$ was also detected in four of them:
OMC3, DR21OH, S140 and W3 IRS5. These detections are important in
that they provide comparisons of another ion with a suitable neutral
species. We also present HCN, HCO$^{+}$ and N$_{2}$H$^{+}$ data
for L1551 IRS5. This brings to eleven the number of objects studied
so far.

As will be shown, the same effects are observed with the less abundant
isotopic species (H$^{13}$CN and H$^{13}$CO$^{+}$); the ion species
generally exhibiting narrower line profiles and significantly suppressed
wings.

Another argument that could be used to counter our assertions is the
importance that chemical differentiation can have on the spectral
appearance of different molecular species. In fact, dramatic abundance
variations between species have been observed in some molecular cores,
as in the case of L1498 \citep{Kuiper 1996}, and one might worry
that species with different spatial distributions would also exhibit
different velocity profiles. But as will be shown, the consistent
success of our specific physical model in predicting the relative
narrowing of the ion lines renders the need for such chemical postulate
less immediate. Moreover, a case can be made against chemical differentiation
being the dominant factor in explaining the relative line widths.
For example, why are the observed ion lines always narrower, not broader
often? Why are the ion lines narrower in cases where the spatial distributions
are shown to be the same? Certainly many factors must be taken into
account in order to explain the differences in the velocity profiles,
but our model's success in answering questions like these forces us
to emphasize the fact that the presence of the magnetic field is most
important and, therefore, should not be neglected.

In the last section of the paper, we discuss recent N$_{2}$H$^{+}$
observations of dense cores in dark clouds \citep{Benson 1998}. As
these cores exhibit, on average, narrow line profiles ($\Delta v\sim0.4$
km/s) and are believed to be primarily thermally supported, they provide
us with a good opportunity to test the assertion made in Paper I that
there should not be any significant differences between the widths
of ion and neutral molecular lines in regions of low turbulence and
void of flows.

\section{Ion velocity behavior.}

It was shown in Paper I that it is possible to calculate the components
of the effective velocity of an ion subjected to a flow of neutral
particles in a region inhabited with a magnetic field. Under the assumptions
that the flow is linear, the plasma is weakly ionized and that all
collisions between the ion and the neutrals are perfectly elastic,
we get for the mean and variance of the ion velocity components:

\begin{eqnarray}
\left\langle \mathbf{v}_{\Vert}\right\rangle  & = & \left\langle \mathbf{v}_{\Vert}^{n}\right\rangle \label{eq:vz}\\
\left\langle \mathbf{v}_{\bot}\right\rangle  & = & \frac{\left\langle \mathbf{v}_{\bot}^{n}\right\rangle +\left\langle \omega_{r}\right\rangle ^{-1}\left[\left\langle \mathbf{v}_{\bot}^{n}\right\rangle \times\left\langle \overrightarrow{\omega_{g}}\right\rangle \right]}{1+\left(\frac{\left\langle \overrightarrow{\omega_{g}}\right\rangle }{\left\langle \omega_{r}\right\rangle }\right)^{2}}\label{eq:vperp}\\
\sigma_{\Vert}^{2} & = & \frac{a\left[\left\langle \left|\mathbf{v}_{\bot}^{n}\right|^{2}\right\rangle -\left\langle \mathbf{v}_{\bot}\right\rangle ^{2}\right]+b\left[\sigma_{\Vert}^{n}\right]^{2}}{\left[\frac{m}{\mu}-1\right]}\label{eq:sigz}\\
\sigma_{\bot}^{2} & = & \frac{g\left[\left\langle \left|\mathbf{v}_{\bot}^{n}\right|^{2}\right\rangle -\left\langle \mathbf{v}_{\bot}\right\rangle ^{2}\right]+h\left[\sigma_{\Vert}^{n}\right]^{2}}{\left[\frac{m}{\mu}-1\right]}\label{eq:sigp}\\
\sigma_{T}^{2} & = & \frac{\left[\left\langle \left|\mathbf{v}_{\bot}^{n}\right|^{2}\right\rangle -\left\langle \mathbf{v}_{\bot}\right\rangle ^{2}\right]+\left[\sigma_{\Vert}^{n}\right]^{2}}{\left[\frac{m}{\mu}-1\right]}\label{eq:sigt}\end{eqnarray}

with

\begin{eqnarray}
\left\langle \overrightarrow{\omega_{g}}\right\rangle  & = & \frac{e\left\langle \mathbf{B}\right\rangle }{mc}\label{eq:gyrofreq}\\
\left\langle \omega_{r}\right\rangle  & \simeq & \frac{\mu}{m}\nu_{c}\label{eq:awrelax}\\
\left[\sigma^{n}\right]^{2} & = & \left\langle \left|\mathbf{v}^{n}\right|^{2}\right\rangle -\left\langle \mathbf{v}^{n}\right\rangle ^{2}\label{eq:ndisp}\end{eqnarray}

$m$ and $\mu$ are the ion mass and the reduced mass. The ion and
neutral flow velocities ($\mathbf{v}$ and $\mathbf{v}^{n}$) were
broken into two components: one parallel to the magnetic field ($\mathbf{v}_{\Vert}$
and $\mathbf{v}_{\Vert}^{n}$) and another ($\mathbf{v}_{\bot}$ and
$\mathbf{v}_{\bot}^{n}$) perpendicular to it. $\left\langle \omega_{r}\right\rangle $,
$\left\langle \overrightarrow{\omega_{g}}\right\rangle $ and $\nu_{c}$
are the relaxation rate, the mean ion gyrofrequency vector and the
(mean) collision rate. 

Under the assumption that the neutral flow consists mainly of molecular
hydrogen and has a mean molecular mass $A_{n}=2.3$, we get $a\simeq0.16$,
$b\simeq0.67$, $g=1-a$ and $h=1-b$ (these are functions with a
weak dependence on the ion and neutral masses, the values given here
apply well to the different ion masses encountered in this paper).

In the {}``strong'' magnetic field limit ($\left\langle \omega_{g}\right\rangle \gg\left\langle \omega_{r}\right\rangle $),
the previous set of equations is simplified by the fact that the ion
gets trapped in the magnetic field, this implies that $\left\langle \mathbf{v}_{\bot}\right\rangle \sim0$
(by equation (\ref{eq:vperp})). This condition is probably easily
met in a typical cloud where the required magnetic field strength,
which scales linearly with the density, can be calculated to be $\left\langle B\right\rangle \ga1\,\mu$G
for a density of $\sim10^{6}$ cm$^{-3}$. 

Upon studying the implications of equations (\ref{eq:vz})-(\ref{eq:ndisp}),
we developed three expectations concerning the differences between
the line profiles of coexistent ion and neutral species:

\begin{itemize}
\item for regions inhabited with a strong enough magnetic field which is,
on average, not aligned with the local flows ($\left\langle \mathbf{v}_{\Vert}^{n}\right\rangle $
is not dominant), we expect ionic lines to exhibit narrower profiles
and a suppression of high velocity wings when compared to neutral
lines since $\left\langle \mathbf{v}_{\bot}\right\rangle \sim0$
\item there should be no significant differences between the line profiles
when the flows and the mean magnetic field are aligned as the ion
velocity is completely determined by $\left\langle \mathbf{v}_{\Vert}\right\rangle $
(or equivalently $\left\langle \mathbf{v}_{\Vert}^{n}\right\rangle $)
\item a \emph{}thermal (or microturbulent) line profile would not show any
manifestation of the presence of the magnetic field since the mean
velocity of the neutral flow is zero.
\end{itemize}
We refer the reader to Paper I for more details.

\subsection{line widths and profiles.\label{sec:lines}}

Equations (\ref{eq:vz})-(\ref{eq:ndisp}) describe the effect the
mean magnetic field has on ions. To derive the line profile for ion
species, one could, in principle, apply this set of equations to each
and every neutral flow present in a region of a given molecular clouds
and project the resulting ion velocities along the line of sight.
The sum of all these contributions would give us the observed line
profile.

Even though we don't have such a detailed knowledge of the dynamics
of molecular clouds, we can still apply this procedure to the study
of simple cases. More precisely, we will concentrate on a given position
in a cloud where all the neutral flows are such that they have i)
azimuthal symmetry about the axis defined by the direction of the
mean magnetic field and ii) reflection symmetry across the plane perpendicular
to the same axis. For example, a bipolar outflow would fit this geometry.
With these restrictions, we get for the observed ion line profiles:

\begin{eqnarray}
\left\langle v_{obs}\right\rangle  & = & 0\label{eq:vobs}\\
\sigma_{obs}^{2} & = & \sum_{k}C^{k}\left[\left\langle \left|\mathbf{v}_{\Vert}^{k}\right|^{2}\right\rangle \cos^{2}\left(\alpha\right)+\frac{1}{2}\left\langle \left|\mathbf{v}_{\bot}^{k}\right|^{2}\right\rangle \sin^{2}\left(\alpha\right)\right]\label{eq:v2obs}\end{eqnarray}

where $\alpha$ is the angle between the direction of the mean magnetic
field and the line of sight. The summation runs over every flow contained
in any given quadrant of any plane which is perpendicular to the plane
of reflection symmetry and which also contains the axis of symmetry.
$C^{k}$ is the weight associated with the flow $k$, which presumably
scales with the ion density (if the line is optically thin). An equivalent
set of equations applies equally well to any coexistent neutral species
(which we assume to exist in proportions similar to that of the ion),
provided we replace $v_{obs}$ by $v_{obs}^{n}$, $\sigma_{obs}$
by $\sigma_{obs}^{n}$ and $\mathbf{v}^{k}$ by $\left[\mathbf{v}^{n}\right]^{k}$.
Using (\ref{eq:vz})-(\ref{eq:ndisp}), one can easily verify that
these two sets of equations (for neutrals and ions) are identical
when the mean magnetic field is negligible ($\left\langle \omega_{g}\right\rangle \ll\left\langle \omega_{r}\right\rangle $).

Under the assumptions that there is no intrinsic dispersion in the
neutral flows and that the mean magnetic field is strong ($\left\langle \omega_{g}\right\rangle \gg\left\langle \omega_{r}\right\rangle $),
we get for the neutral and ion line widths:

\begin{eqnarray}
\left[\sigma_{obs}^{n}\right]^{2} & = & \sum_{k}C^{k}\left\langle \left[\mathbf{v}^{n}\right]^{k}\right\rangle ^{2}\left[\left\langle \cos\left(\theta^{k}\right)\right\rangle ^{2}\cos^{2}\left(\alpha\right)+\frac{1}{2}\left\langle \sin\left(\theta^{k}\right)\right\rangle ^{2}\sin^{2}\left(\alpha\right)\right]\label{eq:v2n}\\
\sigma_{obs}^{2} & \simeq & \sum_{k}C^{k}\left\langle \left[\mathbf{v}^{n}\right]^{k}\right\rangle ^{2}\left[\rule{0in}{5ex}\left\langle \cos\left(\theta^{k}\right)\right\rangle ^{2}\cos^{2}\left(\alpha\right)\right.\nonumber \\
 &  & \,\,\,\,\,\,\,\,\,\,\,\,\left.+\frac{\left\langle \sin\left(\theta^{k}\right)\right\rangle ^{2}}{\left[\frac{m}{\mu}-1\right]}\left[a\cos^{2}\left(\alpha\right)+\frac{g}{2}\sin^{2}\left(\alpha\right)\right]\right]\label{eq:v2ion}\end{eqnarray}

where $\theta^{k}$ is the angle between flow $k$ and the axis of
symmetry. From equations (\ref{eq:v2n})-(\ref{eq:v2ion}) one can
see that if the flows are aligned with the mean magnetic field ($\theta^{k}\simeq0$),
then neutrals and ions will have similar line widths. On the other
hand, when the flows are perpendicular to the field ($\theta^{k}\simeq\frac{\pi}{2}$)
the ions have a much narrower profile which scales with their molecular
mass as $\left[\frac{m}{\mu}-1\right]^{-\frac{1}{2}}$.

We can further consider the average width one would expect for both
kinds of molecular species when a sample of objects is considered.
This can be achieved by averaging equations (\ref{eq:v2n})-(\ref{eq:v2ion})
over the angle $\alpha$ (assuming the sources in the sample to be
more or less similar). Assuming that there is no privileged direction
in space for the mean magnetic field, we get:

\begin{eqnarray}
\left\langle \left[\sigma_{obs}^{n}\right]^{2}\right\rangle  & = & \frac{1}{3}\sum_{k}C^{k}\left\langle \left[\mathbf{v}^{n}\right]^{k}\right\rangle ^{2}\label{eq:v2nsamp}\\
\left\langle \sigma_{obs}^{2}\right\rangle  & \simeq & \frac{1}{3}\sum_{k}C^{k}\left\langle \left[\mathbf{v}^{n}\right]^{k}\right\rangle ^{2}\left[\left\langle \cos\left(\theta^{k}\right)\right\rangle ^{2}+\frac{\left\langle \sin\left(\theta^{k}\right)\right\rangle ^{2}}{\left[\frac{m}{\mu}-1\right]}\right]\,.\label{eq:v2ionsamp}\end{eqnarray}

A variation of the angle $\alpha$ between the mean magnetic field
and the line of sight can have a significant effect on the width of
the observed ion line profile from a given object (see equation (\ref{eq:v2ion})).
But this is not necessarily true for neutral species, we can see from
equation (\ref{eq:v2n}) that the line width is a lot less sensitive
to variations in $\alpha$ than that of the ion. The same statements
are true for the sample of similar sources (of different orientations)
considered here. We can therefore expect the mean width of a neutral
line to be larger than its variation between sources and make the
following approximation:

\[
\frac{\left\langle \sigma_{obs}^{2}\right\rangle }{\left\langle \left[\sigma_{obs}^{n}\right]^{2}\right\rangle }\simeq\left\langle \frac{\sigma_{obs}^{2}}{\left[\sigma_{obs}^{n}\right]^{2}}\right\rangle \,.\]

The term on the right hand side represents the average of the square
of the ratio of the ion to neutral line widths. By computing this
ratio, one can get an idea of the propensity shown by the flows and
the mean magnetic field for alignment with each other. 

Going back to equations (\ref{eq:v2nsamp})-(\ref{eq:v2ionsamp}),
we once again see that if the flows are parallel to the field ($\theta^{k}\simeq0$),
the mean ratio of ion to neutral line widths is close to unity. When
the flows are perpendicular ($\theta^{k}\simeq\frac{\pi}{2}$) to
the field, the square of this ratio is proportional to $\left[\frac{m}{\mu}-1\right]^{-1}$.
On the other hand, if the flows don't have any preferred direction
and are randomly oriented (with a uniform distribution) we then find:

\begin{equation}
\left\langle \frac{\sigma_{obs}^{2}}{\left[\sigma_{obs}^{n}\right]^{2}}\right\rangle \simeq\frac{1}{3}\left[1+\frac{2}{\frac{m}{\mu}-1}\right]\label{eq:ratio}\end{equation}

which equals 0.38 (0.37) for an ion of molecular mass $A_{i}=30$
(45).

\section{Observations of H$^{13}$CN, H$^{13}$CO$^{+}$ and HCS$^{+}$.}

In this section we complement our previous observations of HCN, HCO$^{+}$
and N$_{2}$H$^{+}$ in OMC1, OMC2, OMC3 and DR21OH with observations
of H$^{13}$CN, H$^{13}$CO$^{+}$ and HCS$^{+}$ on a sample of ten
objects. These observations are important in that they allow us to
ensure that the effect reported in Paper I was not caused by excess
saturation of the HCN line profiles. H$^{13}$CN, H$^{13}$CO$^{+}$
and HCS$^{+}$ are substantially less abundant than HCN and HCO$^{+}$,
so their emission lines are more likely to be optically thin. Another
fundamental reason for the importance of these further observations
is that the effects of the magnetic field should be felt by all species
of ions. It is therefore desirable to test our concept with observations
from as many ion species as possible.

\begin{figure}
\notetoeditor{the four EPS files should appear in the figure fig:spectra as shown with the commands below}

\begin{center}\epsscale{0.9}\plottwo{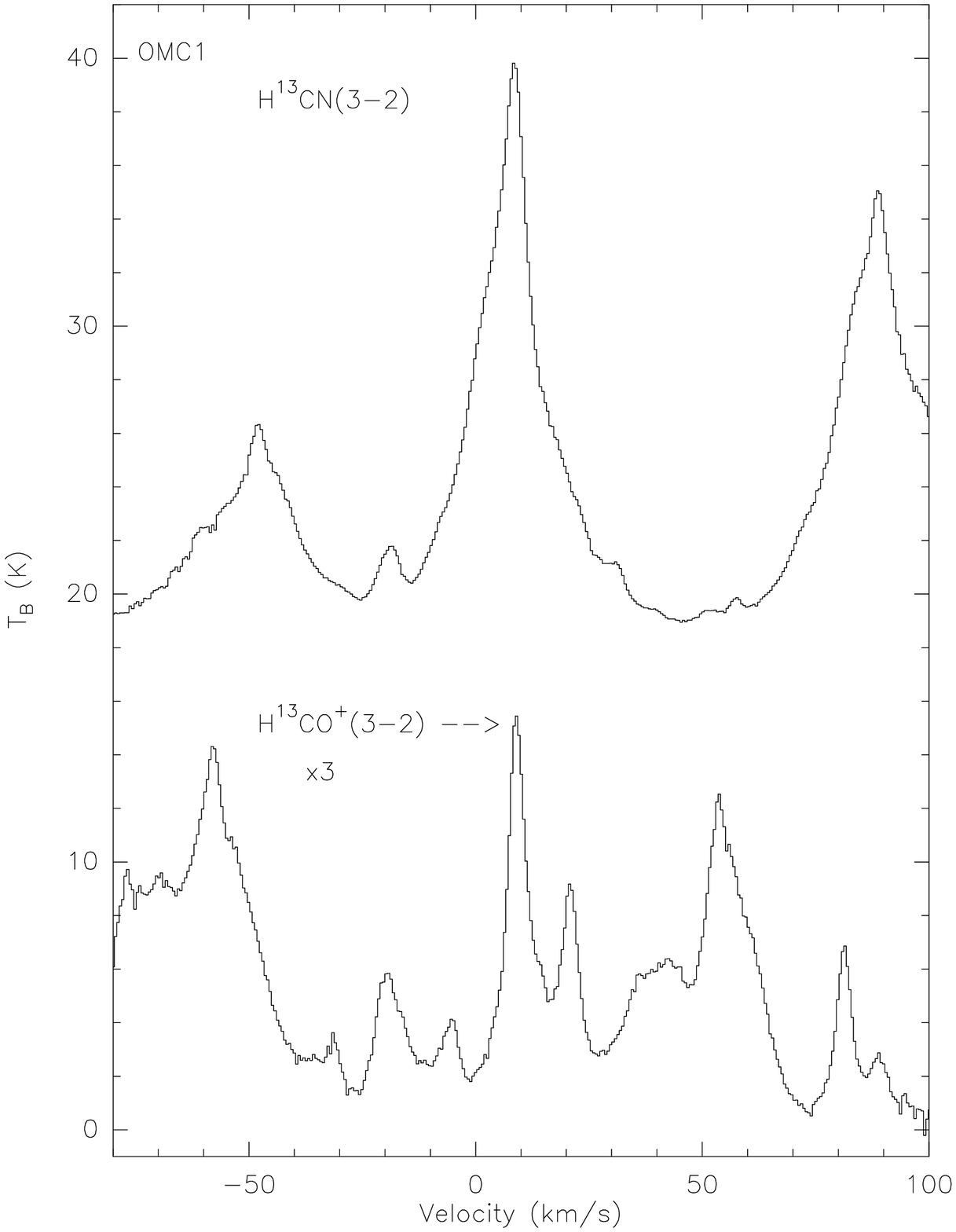}{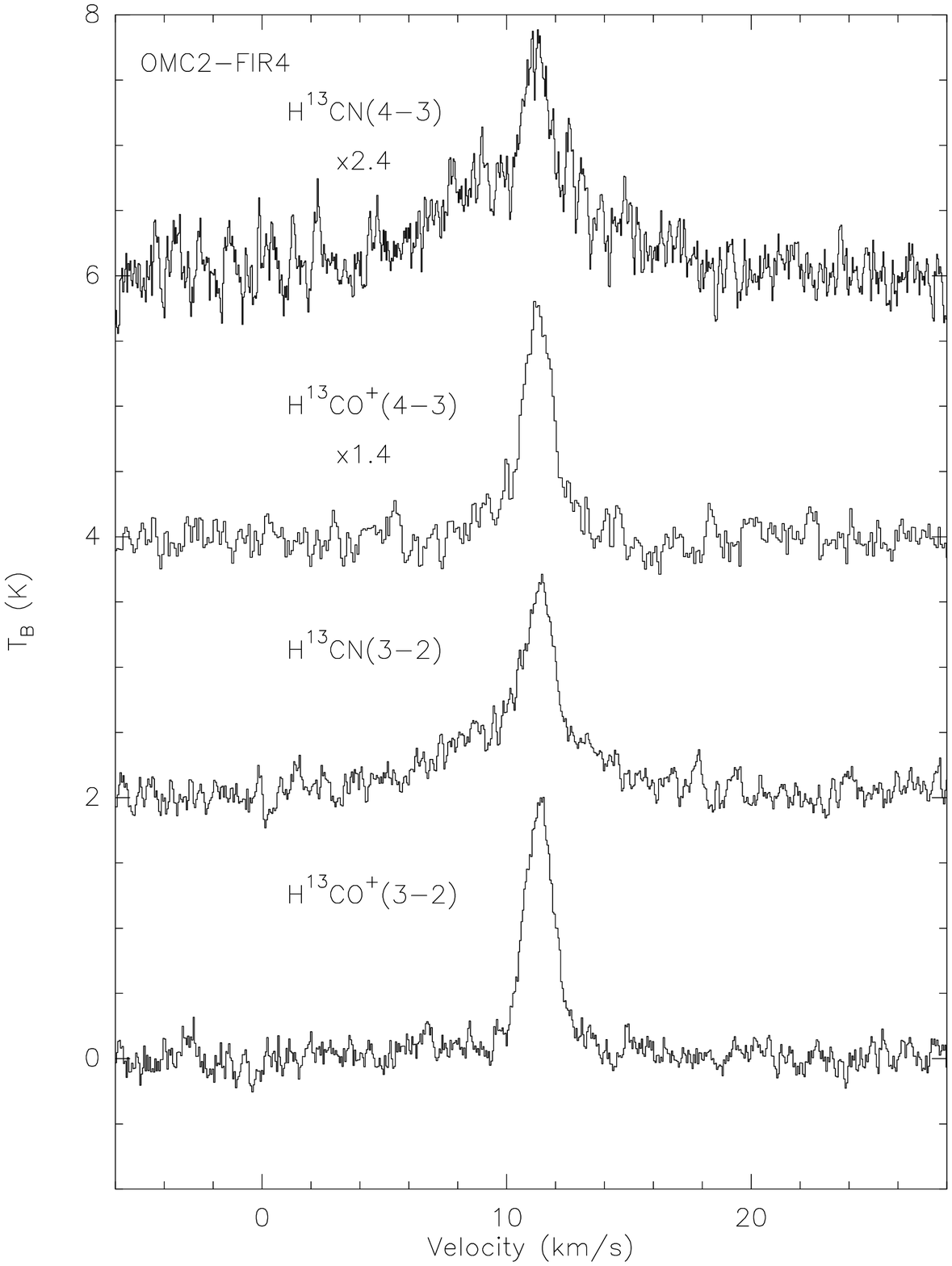}\end{center}

\begin{center}\epsscale{0.9}\plottwo{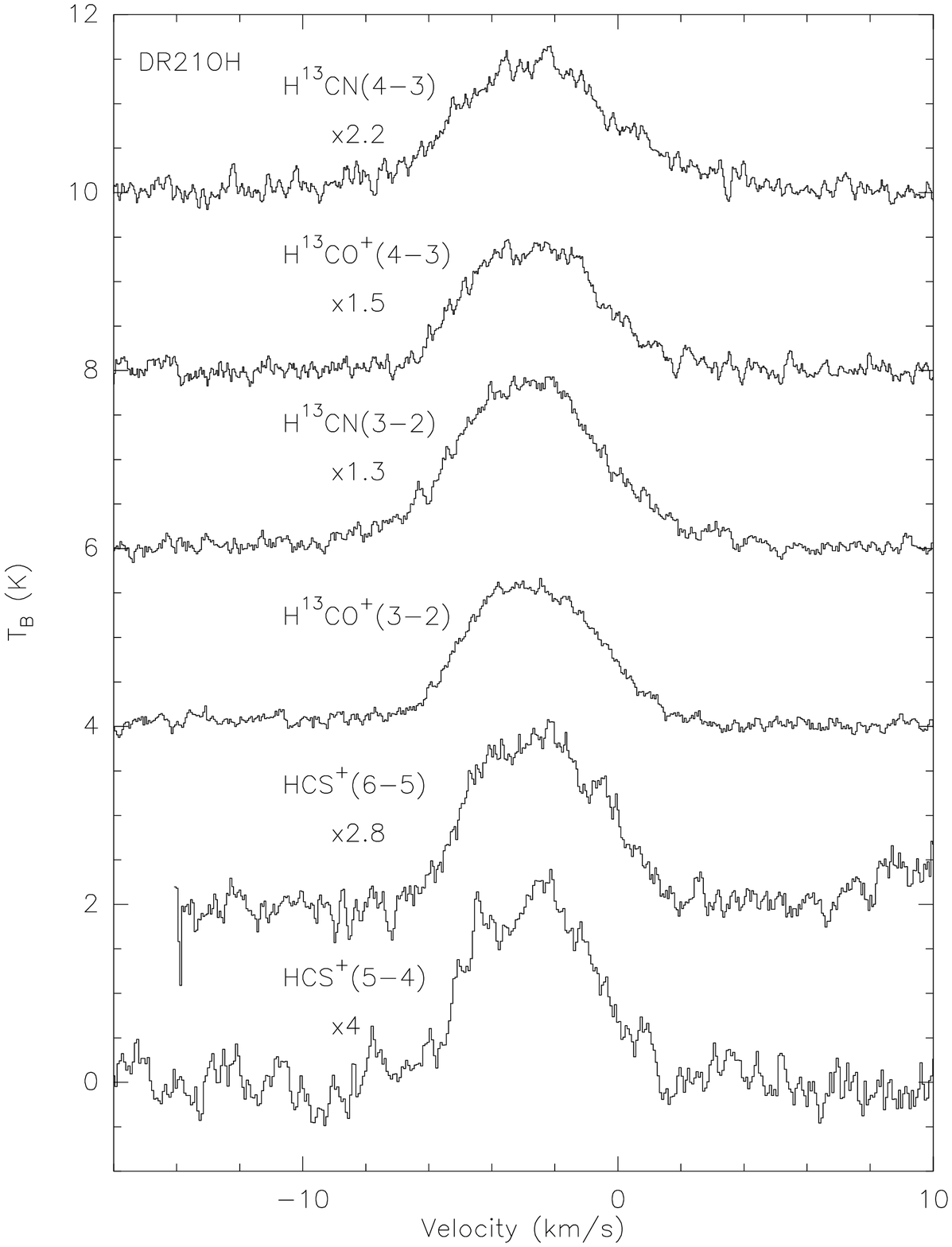}{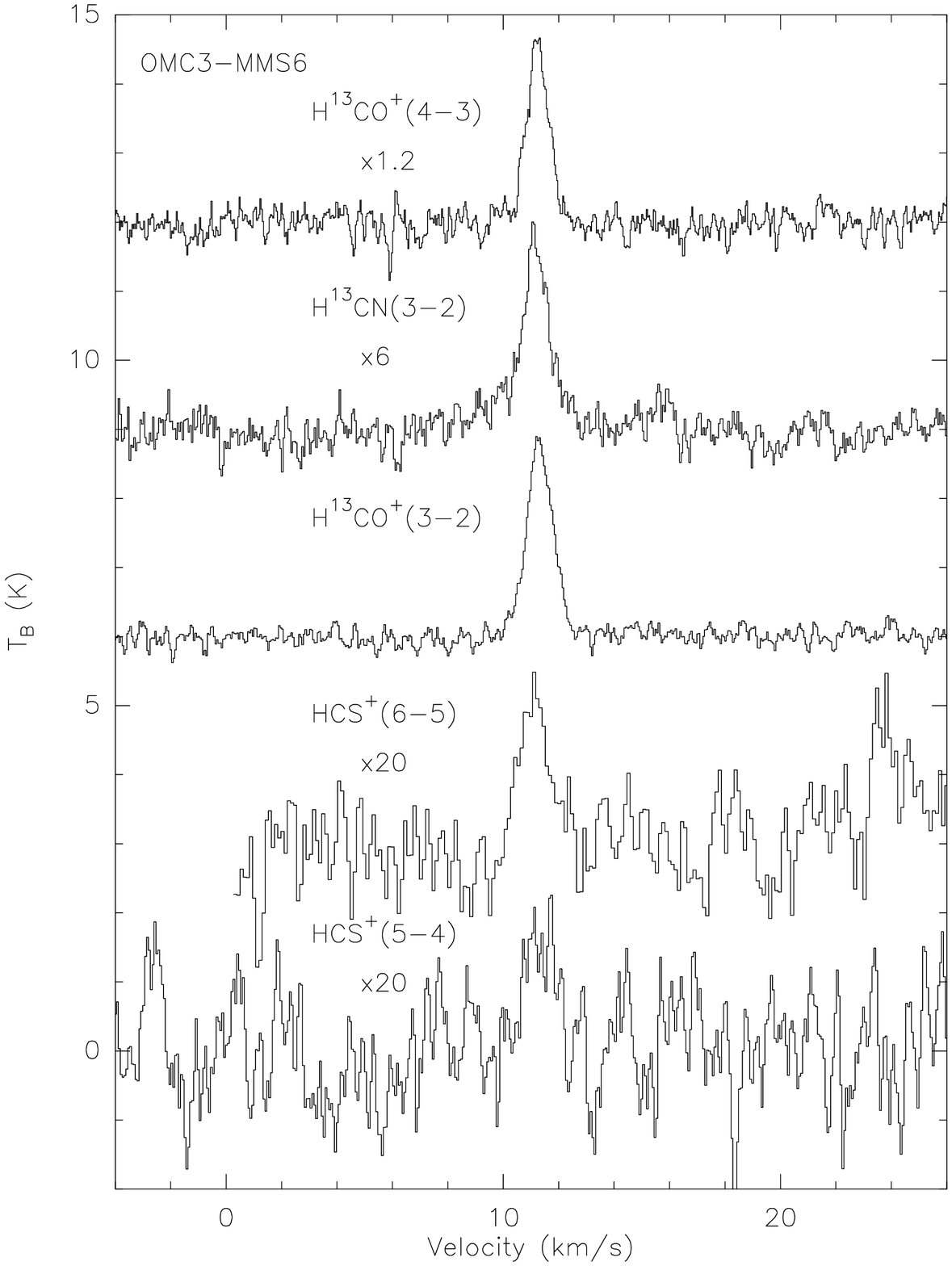}\end{center}

\caption{\label{fig:spectra}H$^{13}$CN , H$^{13}$CO$^{+}$ and HCS$^{+}$
(for DR21OH and OMC3-MMS6) observations at the center position of
(clockwise starting from top left): OMC1, OMC2-FIR4, OMC3-MMS6 and
DR21OH. }
\end{figure}

\begin{figure}
\notetoeditor{the four EPS files should appear in the figure fig:spectra2 as shown with the commands below}

\begin{center}\epsscale{0.9}\plottwo{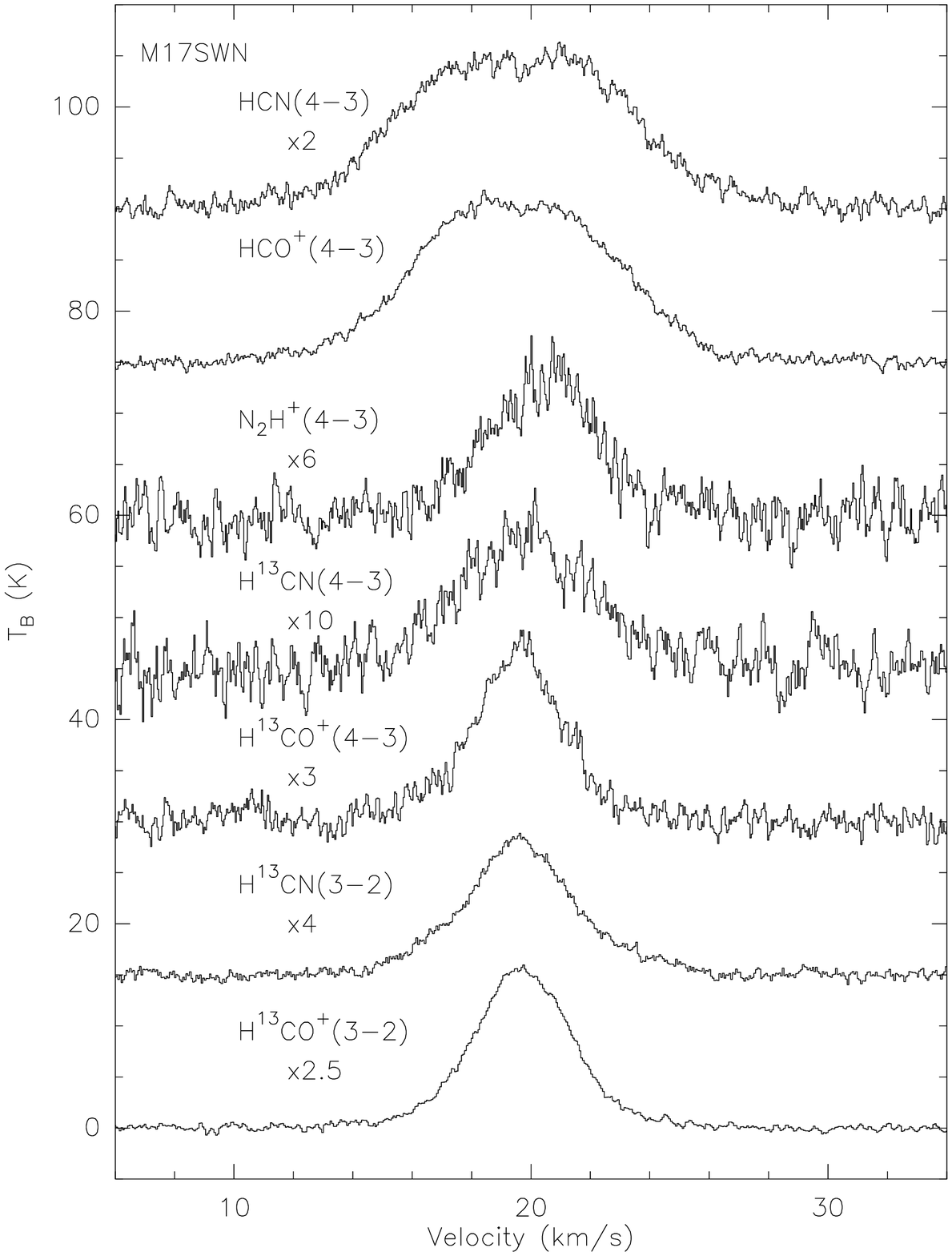}{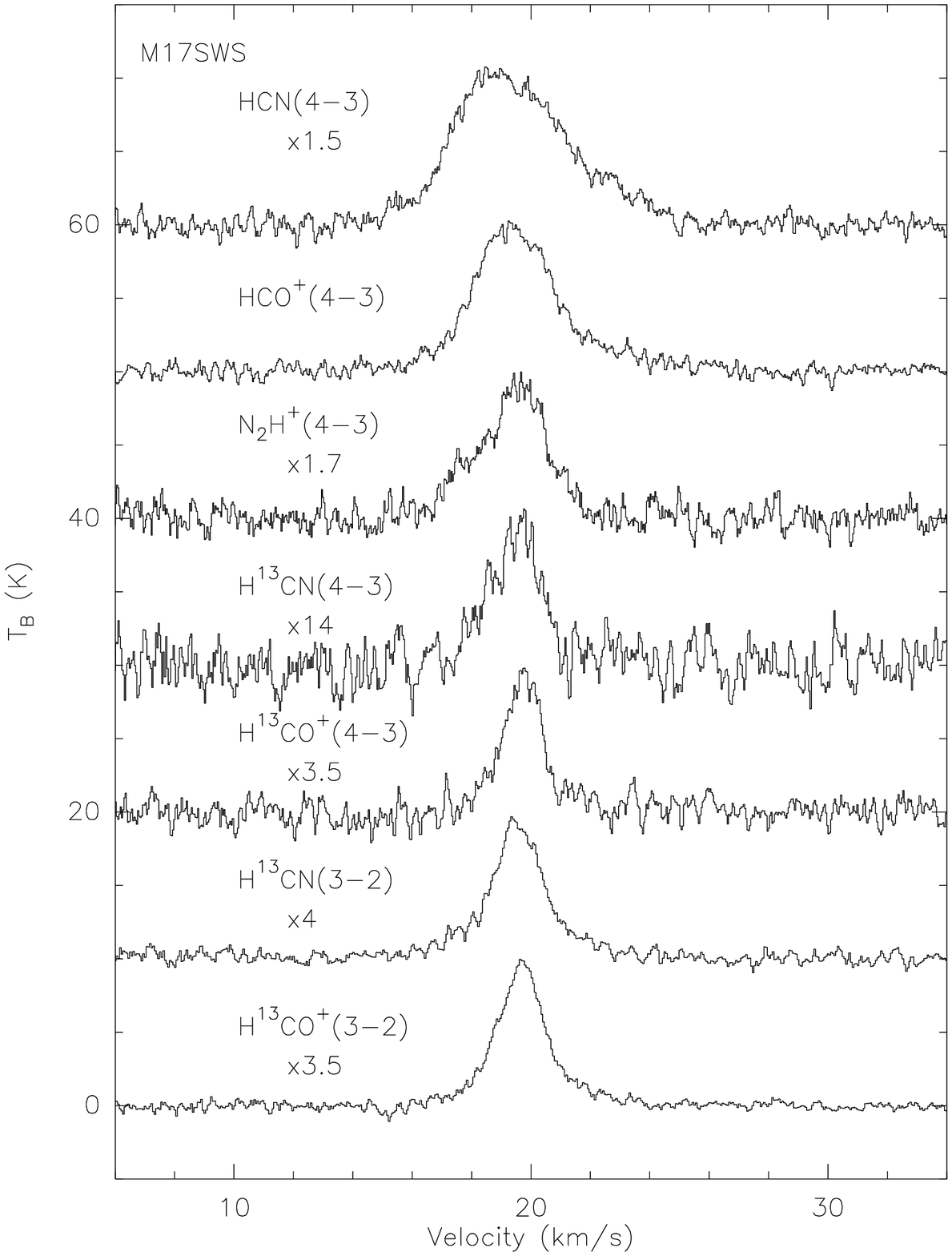}\end{center}

\begin{center}\epsscale{0.9}\plottwo{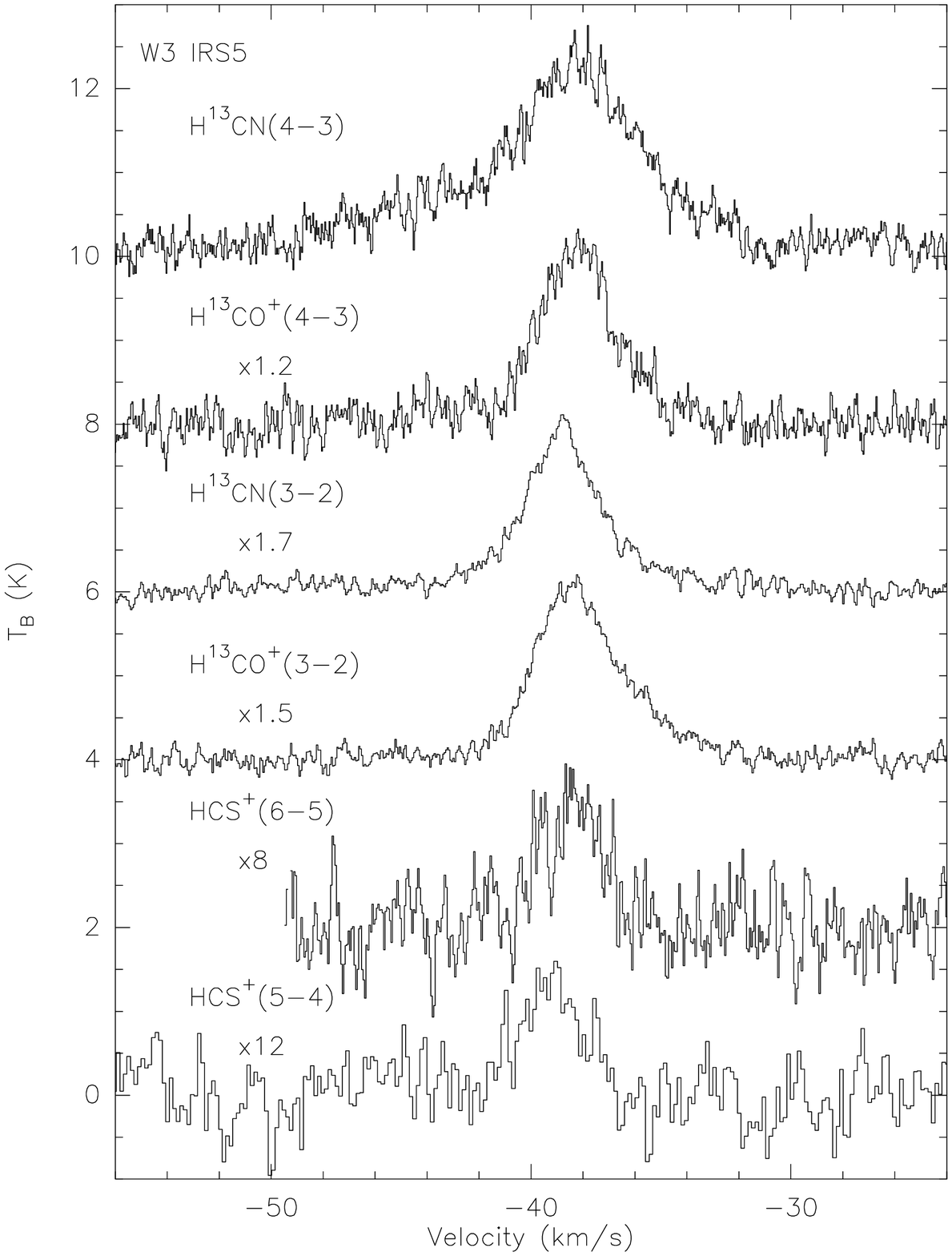}{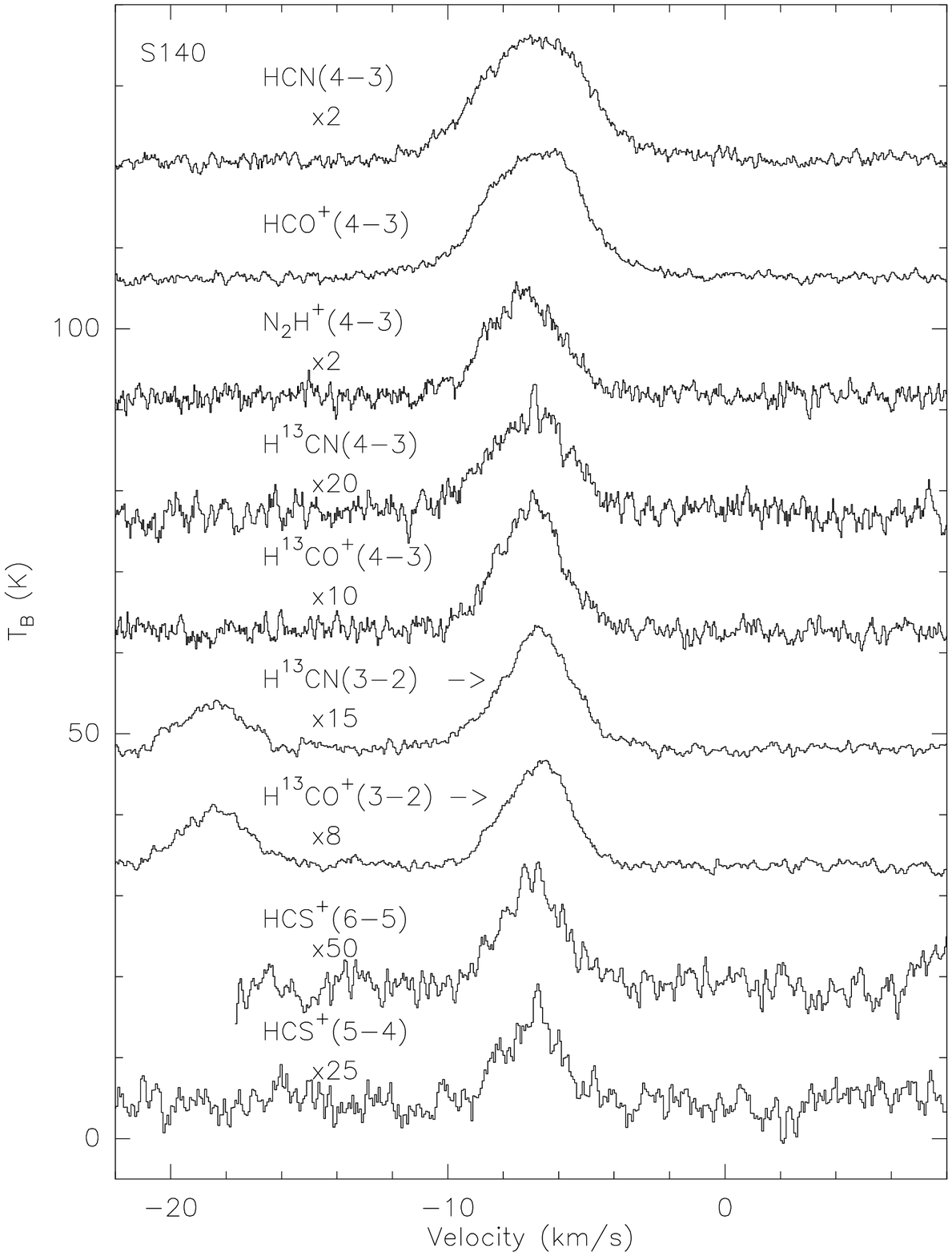}\end{center}

\caption{\label{fig:spectra2}H$^{13}$CN , H$^{13}$CO$^{+}$ and HCS$^{+}$
(for S140 and W3 IRS5) observations at the center position of (clockwise
starting from top left): M17SWN, M17SWS, S140 and W3 IRS5. }
\end{figure}

\begin{figure}
\notetoeditor{the two EPS files should appear in the figure fig:spectra3 as shown with the commands below}

\begin{center}\plottwo{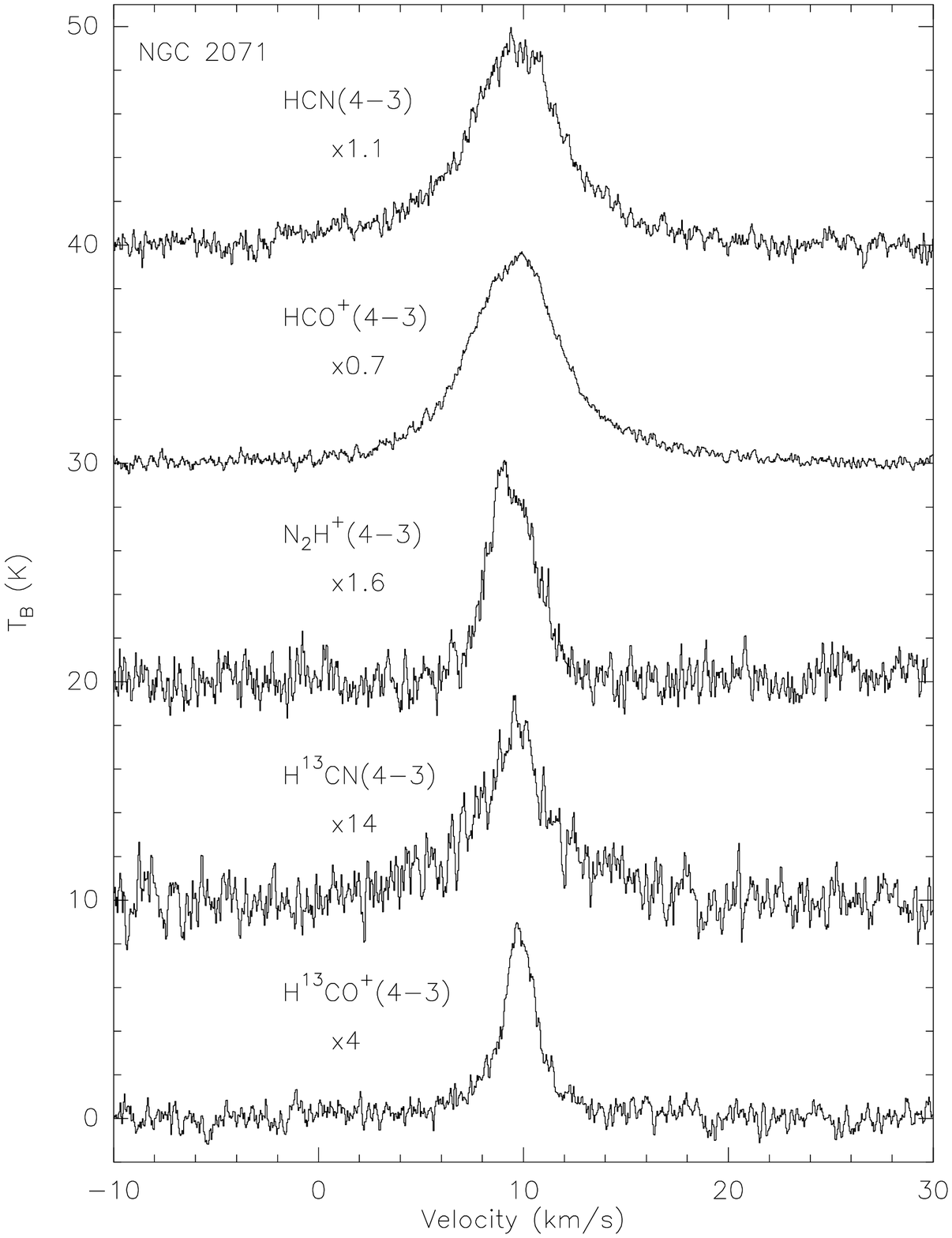}{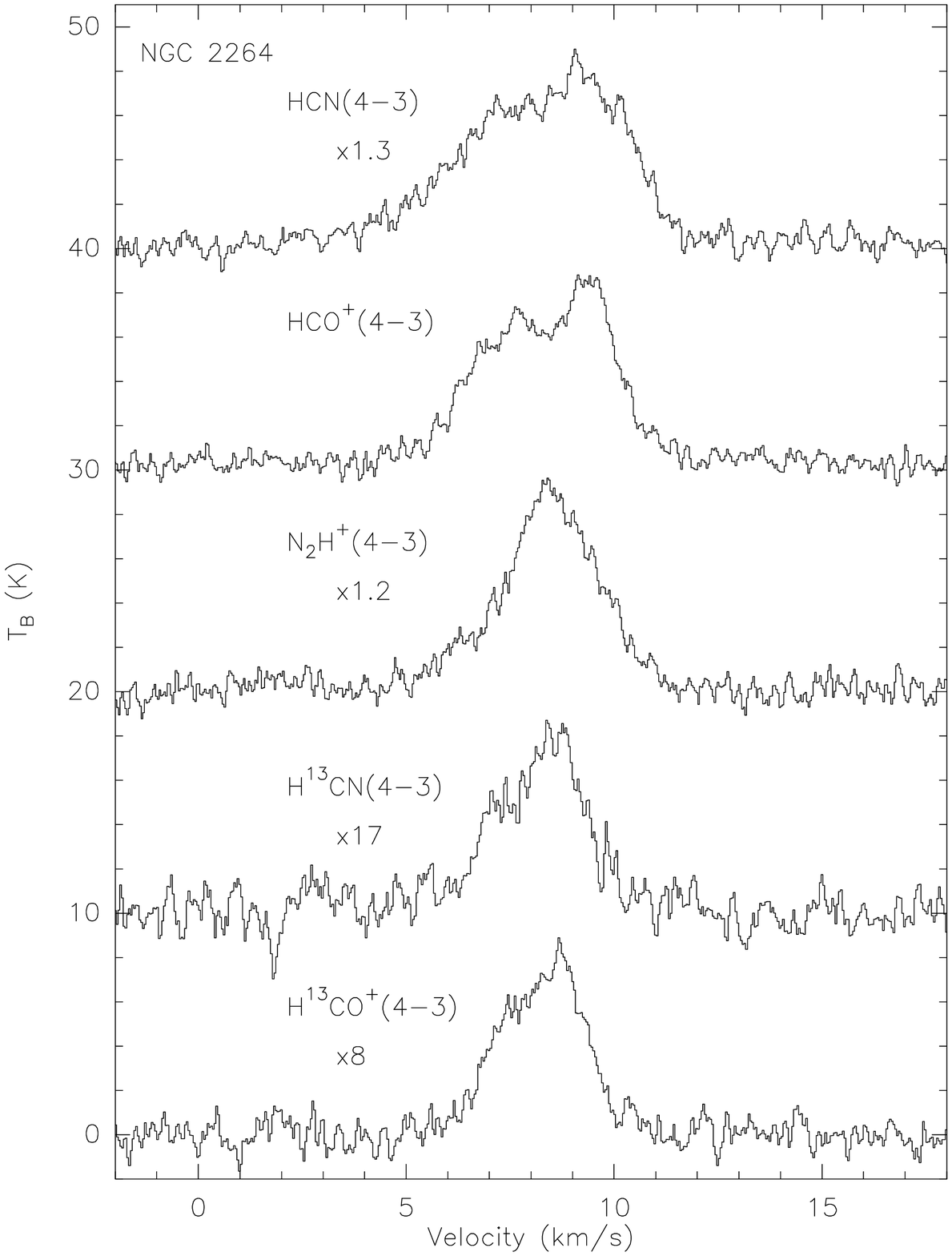}\end{center}

\caption{\label{fig:spectra3}HCN, HCO$^{+}$, N$_{2}$H$^{+}$, H$^{13}$CN
and H$^{13}$CO$^{+}$ observations at the center position of NGC
2071 (left) and NGC 2264 (right). }
\end{figure}

\begin{figure}
\notetoeditor{the four EPS files should appear in the figure fig:maps as shown with the commands below}

\resizebox*{16cm}{7.5cm}{\rotatebox{270}{\includegraphics{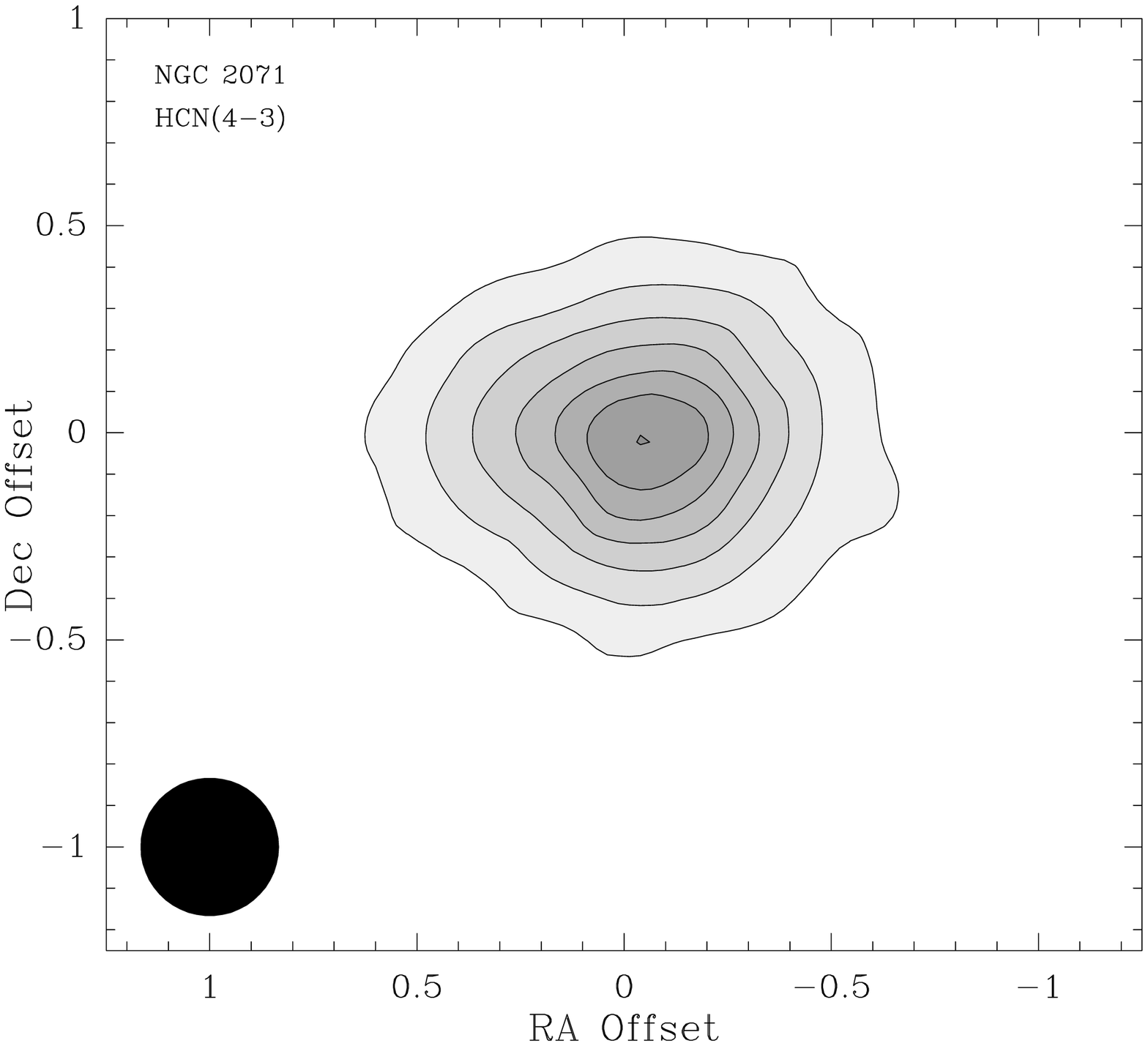}}\hspace{0.5cm}\rotatebox{270}{\includegraphics{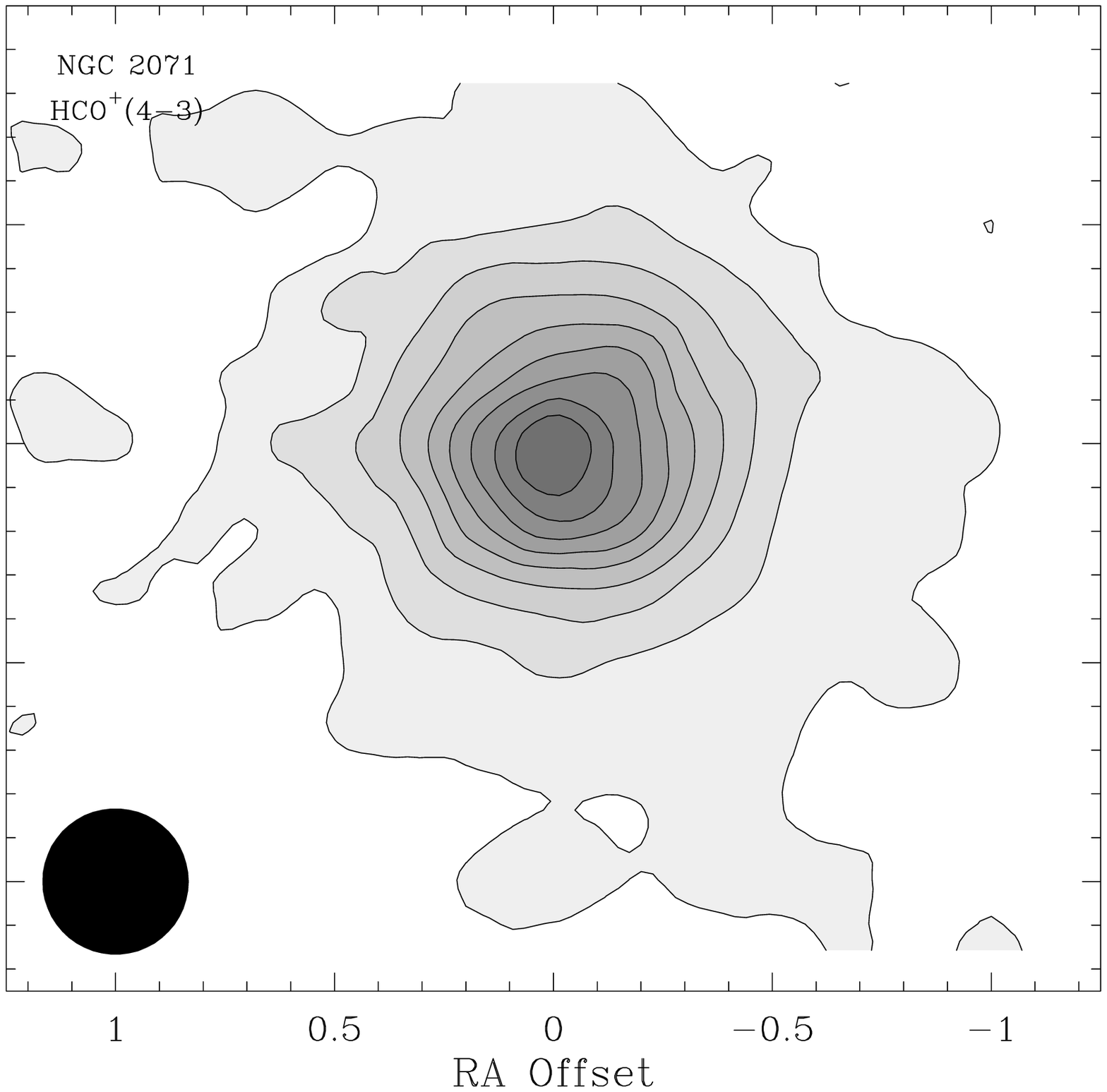}}}

\vspace{0.5cm}

\resizebox*{16cm}{7cm}{\rotatebox{270}{\includegraphics{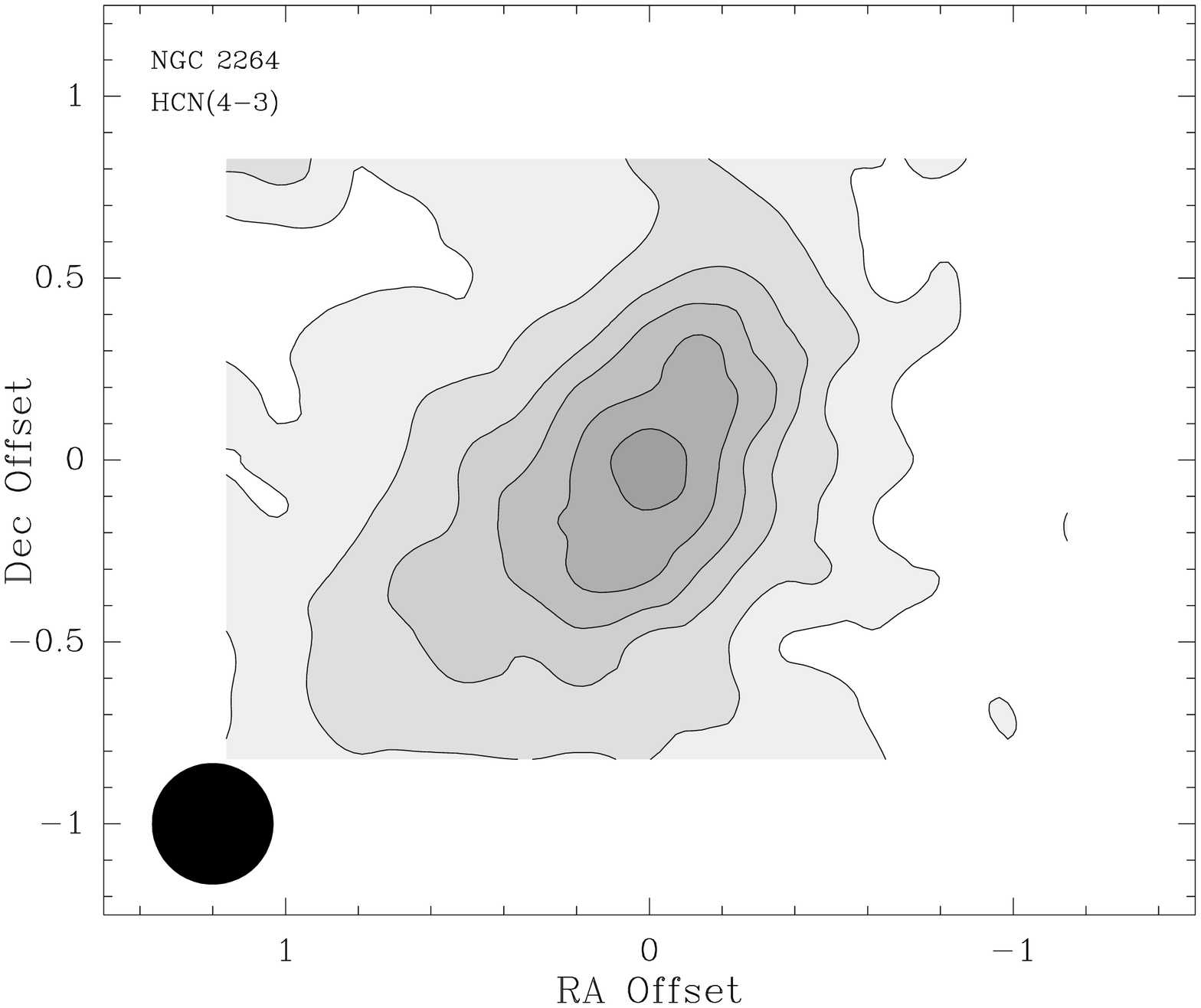}}\hspace{0.5cm}\rotatebox{270}{\includegraphics{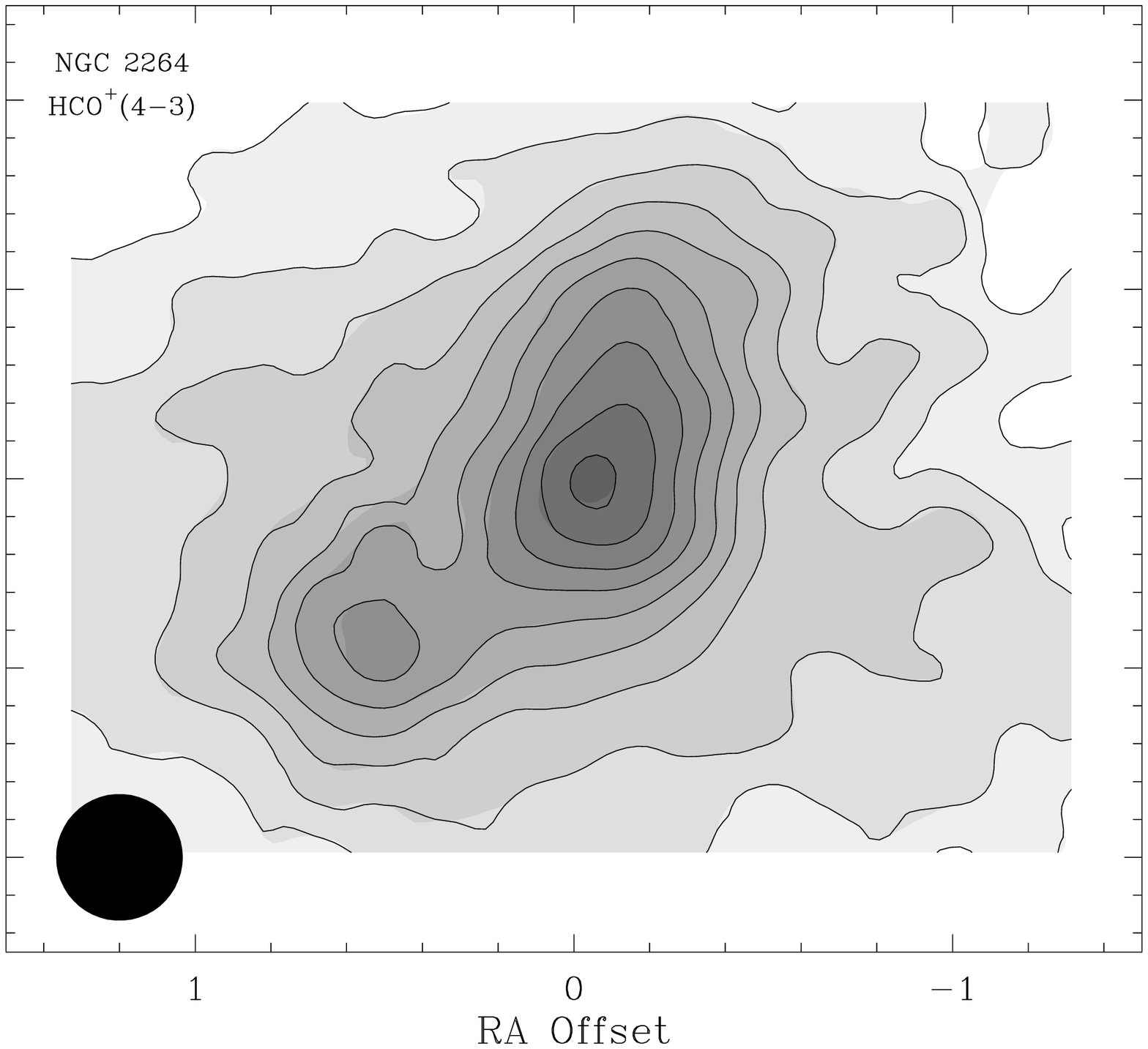}}}

\vspace{1cm}

\caption{\label{fig:maps}HCN and HCO$^{+}$ ($J\rightarrow4-3$) maps of
NGC 2071 (top) and NGC 2264 (bottom). Although the ion spatial distributions
are somewhat more extended, the two peaks are well aligned and the
HCN and HCO$^{+}$ systematic velocities agree (from figure \ref{fig:spectra3}).
The maps for NGC 2071 (NGC 2264) have the same lowest contour level
of 5 K$\cdot$km/s (2.5 K$\cdot$km/s) and the following levels increase
linearly with an interval of the same amount. The maps' grid spacing
of $10\arcsec$ is half of the beam width (shown in the lower left
corners). The pointing accuracy is better than $\sim5\arcsec$.}
\end{figure}

We present the spectra in figures \ref{fig:spectra}, \ref{fig:spectra2}
and \ref{fig:spectra3}. All observations were obtained with the 200-300
GHz and 300-400 GHz receivers at the CSO over a total of fifteen nights
during the months of October, November and December 1999. The spectra
were calibrated using scans made on planets available during this
period (Mars, Jupiter and Saturn). Telescope efficiencies were calculated
to be $\sim70$ \% for the 200-300 GHz receiver (beam width of $\sim32\arcsec$)
and $\sim60$ \% for the 300-400 GHz receiver (beam width of $\sim20\arcsec$).
The maps of NGC 2071 and NGC 2264 in HCN and HCO$^{+}$ ($J\rightarrow4-3$)
presented in figure \ref{fig:maps} were obtained with the 300-400
GHz receiver in April 1999.

As can be seen, all objects show the line profiles of the ion species
as being narrower than those of the neutral species. This is most
evident in OMC1, OMC2-FIR4 and W3 IRS5. A suppression of the high
velocity wings is the most obvious feature in the case of DR21OH. 

Upon closer inspection of the spectra obtained for OMC2-FIR4, we note
that the neutral species exhibit profiles that can easily be separated
into two components: one narrow line with a FWHM $\sim1$ km/s that
we can associate with the quiescent part of the cloud and a broad
feature with a FWHM $\sim8$ km/s which we assume to be related to
(out)flows. On the other hand, the flow component is virtually non-existent
in the ion spectra. From this, we can conclude that there is probably
a significant misalignment between the direction of the flows and
that of the mean magnetic field for this object.

In the case of DR21OH, it is worth noting the striking difference
between this set of spectra and the one obtained for the main isotopes
(see figure 2 of Paper I). Indeed, a double peaked profile was the
most obvious feature in the HCN and HCO$^{+}$ spectra. This was also
observed in CN observations by \citet{Crutcher et al. 1999} which
lead them to identify the peaks as representing two different velocity
components and assign two different magnetic field values as measured
with the Zeeman effect. But as pointed out before, these molecular
species are certainly abundant enough that we should not be surprised
if they exhibited strongly self-absorbed line profiles. This effect
is well known and can lead to an underestimation of the kinetic temperature
of molecular clouds as described, for instance, by \citet{Phillips 1981}.
The fact that the less abundant isotopes do not exhibit the same double
peaked profile suggests that this is what is at play in the case of
DR21OH and that we are really dealing with a single velocity component.

\placefigure{fig:spectra}

\placefigure{fig:spectra2}

\placefigure{fig:spectra3}

We present in table \ref{ta:widths} a comparison of the line widths
(more precisely their standard deviations $\sigma_{v}$) for the different
species discussed in this paper, as well as for those studied in Paper
I, for our sample of molecular clouds. The widths were measured after
the lines were modeled with a multi-Gaussian profile. We have separated
the data into four different groups within which the comparison between
ion and neutral species should be made. The separations occur naturally
and are based on two criteria: i) the perceived optical depth of the
lines and ii) the telescope beam applicable to different sets of observations
(for example, for H$^{13}$CN (and H$^{13}$CO$^{+}$) in the $J\rightarrow4-3$
transition, the beam size is $\sim20\arcsec$ whereas it is $\sim32\arcsec$
for the lower $J\rightarrow3-2$ transition). We have also included
two detections of H$_{3}$O$^{+}$ made on W3 IRS5 by \citet{Phillips et al. 1992},
these are important since they allow us to test our theory on a different
ion species which has a significantly lower molecular mass ($A_{i}=19$).

We have, whenever possible, verified that the opacity of the lines
of the H$^{13}$CN, H$^{13}$CO$^{+}$, HCS$^{+}$ and H$_{3}$O$^{+}$
species are indeed thin. This was accomplished using the ratio of
the line temperature between two different transitions of a given
species (assuming LTE and unity beam filling factors \citep{Emerson 1996}).
For example, in the case of the H$^{13}$CO$^{+}$ spectra for OMC2-FIR4
we have line temperatures of 1.9 K and and 1.3 K for the $J\rightarrow3-2$
and $J\rightarrow4-3$ transitions respectively. From their ratio
and the assumptions made above, we calculated an excitation temperature
of $\sim11$ K (assumed to be the same for both transitions) which
then in turn allows us to obtain opacities of $\sim0.17$ and $\sim0.12$
for the $J\rightarrow3-2$ and $J\rightarrow4-3$ transitions respectively.
It should be kept in mind that LTE may not be appropriate for these
transitions/sources. Furthermore, the values thus obtained are quite
sensitive to the beam filling factors which are probably different
for lines observed with different beam sizes. However, we are not
seeking a precise value for the opacities, but merely a reassurance
that these lines are thin, which this technique provides. 

In all cases, the lines were found to be optically thin with the ions
generally more opaque than the neutrals, the highest opacity calculated
was $\sim0.3$. One possible exception is the H$^{13}$CN $J\rightarrow3-2$
line for OMC1, where a comparison with the same transition of the
main isotope gives an opacity of $\sim0.6$. At any rate, it appears
that we can safely assume that differences in opacity between ion
and neutral lines won't bring any ambiguity in studying the effects
of the magnetic field (see section \ref{sec:intro}). We can therefore
be confident that our widths comparisons are justified.

Upon examination of table \ref{ta:widths}, we note that in all cases
the ion lines exhibit smaller widths ($\sigma_{v}$) than the comparable
neutral lines. As was stated in Paper I, this difference in line width
could also be partly due to other factors (sampling effect, chemical
differentiation, \ldots{}). But to bring support to the hypothesis
that HCN and HCO$^{+}$ are suitable candidate for our study, we discussed
observational evidence from the extensive work of \citet{Ungerechts 1997},
more precisely the fact that their maps of the region surrounding
OMC1 for these two molecular species show similar spatial distributions.
We, in turn, provided our own maps of OMC2-FIR4 for the same species
which also show resemblance in the distributions as well as a good
alignment of their respective peaks. We add to the evidence with figure
\ref{fig:maps} where we again present HCN and HCO$^{+}$ maps in
the $J\rightarrow4-3$ transitions, but this time for NGC 2071 and
NGC 2264. The same comments apply for those maps: although the ion
spatial distributions are somewhat more extended, the two peaks are
well aligned. And from figure \ref{fig:spectra3}, we can also attest
on the agreement of the systematic velocities for the two species. 

\placefigure{fig:maps}

If all is well with HCN and HCO$^{+}$, it appears reasonable to assume
that the same is true for their less abundant isotopic counterparts
H$^{13}$CN and H$^{13}$CO$^{+}$. But we should point out that such
comments are probably not true for N$_{2}$H$^{+}$ when compared
to HCN. It is well known \citep{Bachiller 1996, Bachiller 1997} that
N$_{2}$H$^{+}$ is only observed in the colder condensations of the
gas whereas HCN often takes part in outflows, this would automatically
result in the former exhibiting narrower line profiles. Our conclusions
will, therefore, not be based on comparisons from the data obtained
for these two species. 

This being said, we interpret the fact that we consistently detect
narrower ion line profiles over a sizable sample of molecular clouds
for many different ion species as strong evidence in favor of our
theory. The ratio of ion to neutral line widths range anywhere from
$\sim0.2$ to $\sim0.9$ (see table \ref{ta:ratios}), which is not
unexpected considering the analysis made in section \ref{sec:lines}.
We then showed that ion line widths are strongly dependent on the
alignment of the neutral flows with the mean magnetic field (see equation
(\ref{eq:v2ion})) and are therefore liable to vary from one object
to the next. The largest value for this ratio is $\simeq0.93$ in
the case of NGC 2071, implying a good alignment between the field
and the flows. Interestingly, this object is known to exhibit strongly
collimated jets \citep[and references therein]{Girart et al. 1999},
this suggests the possibility that these are closely linked to the
direction of the mean magnetic field. This aspect will be studied
in detail in a subsequent paper.

As a final comment in this section, we point out that the effect of
the magnetic field on ion line profiles appears to be roughly similar
in importance whether the lines observed are optically thick or thin.
This can seen from table \ref{ta:ratios} where we show the ion to
neutral line width ratios for our sample of objects.

\placetable{ta:ratios}

\placetable{ta:widths}

\subsection{Statistics of the ratios of line widths.}

Following our discussion at the end of section \ref{sec:lines}, it
is interesting to average the square of the ratio of the ion to neutral
line widths over our sample and compare the results with equation
(\ref{eq:ratio}). We find:

\begin{eqnarray}
\left\langle \frac{\sigma_{\mbox{\scriptsize H}^{13}\mbox{\scriptsize CO}^{+}}^{2}}{\sigma_{\mbox{\scriptsize H}^{13}\mbox{\scriptsize CN}}^{2}}\right\rangle  & = & 0.42\label{eq:H13CO+}\\
\left\langle \frac{\sigma_{\mbox{\scriptsize HCS}^{+}}^{2}}{\sigma_{\mbox{\scriptsize H}^{13}\mbox{\scriptsize CN}}^{2}}\right\rangle  & = & 0.32\label{eq:HCS+}\end{eqnarray}

which we interpret as meaning that, although this ratio can vary significantly
from one object to the next, there is, on average, no obvious propensity
for the alignment between the flows and the mean magnetic field in
high density molecular clouds such as those studied here. For we know,
from the aforementioned discussion, that if the neutral flows don't
have any preferred orientation in relation to the magnetic field,
these ratios would be 0.38 for H$^{13}$CO$^{+}$ and 0.37 for HCS$^{+}$.
It should, however, be kept in mind that our sample of objects is
not entirely composed of (more or less) similar objects, as was assumed
when we derived equation (\ref{eq:ratio}).

\section{Regions of low turbulence.}

All the objects discussed so far present a fair amount of turbulence
as can be attested by a comparison of the widths of the observed line
profiles with their expected thermal width. The former being many
times broader than the latter. In fact, as was mentioned earlier we
would only expect to observe narrower ion lines in situations where
this is the case. According to equations (\ref{eq:vz})-(\ref{eq:ndisp})
and (\ref{eq:v2ion}), we can identify three scenarios for which one
should not expect any significant difference between the line widths
of ion and neutral species:

\begin{itemize}
\item the value of the mean magnetic field is such that $\left\langle \omega_{g}\right\rangle \ll\left\langle \omega_{r}\right\rangle $
\item the mean magnetic field is aligned with the flow(s)
\item the region under study has little or no (macro)turbulence.
\end{itemize}
In what follows, we will investigate some observational evidence that,
we will argue, show cases where the last scenario is at work.

\subsection{Dark clouds with dense cores.}

Recently, \citet{Benson 1998} studied, among other things, the correlation
of the velocity and line widths between the spectra of N$_{2}$H$^{+}$
and other neutral species (C$_{3}$H$_{2}$, CCS and NH$_{3}$) on
large samples of dark clouds (with dense cores). Their sample contains
sources which are very different from those that we have been studying
so far in that: 

\begin{itemize}
\item they have much narrower line profiles, the mean line widths (FWHM)
are $0.37$ km/s for N$_{2}$H$^{+}$, $0.46$ km/s for C$_{3}$H$_{2}$,
$0.38$ km/s for CCS and $0.36$ km/s for NH$_{3}$
\item most of the cores are primarily thermally supported, the corresponding
thermal width of a neutral molecule of mean mass is $0.45$ km/s at
$10$ K
\item the densities probed with their observations are roughly two orders
of magnitude lower, $n\ga10^{4}$ cm$^{-3}$ in their case compared
to $n\ga10^{6}$ cm$^{-3}$ in ours.
\end{itemize}
As stated before, we expect that regions of low turbulence such as
these dark cloud cores would not show significant differences between
the width of the line profiles of ion and neutral species. The measured
mean line widths reported above seem to indicate that this is the
case, only the C$_{3}$H$_{2}$ species shows a mean line width noticeably
different from that of the ion species. However, this situation improves
when we limit the comparisons to only those objects which are common
to each sample (i.e., the different sets of N$_{2}$H$^{+}$, CCS
and C$_{3}$H$_{2}$ observations). This group of objects is listed
in table \ref{ta:Benson} along with the measured line widths. With
this restriction, we find that the mean line widths are now 0.36 km/s
for N$_{2}$H$^{+}$, 0.37 km/s for CCS and 0.41 km/s for C$_{3}$H$_{2}$.
More to the point, if we calculate the root mean square value of the
line widths ratios between the different species we get:

\begin{eqnarray*}
\sqrt{\left\langle \frac{\Delta v_{\mbox{\scriptsize N}_{2}\mbox{\scriptsize H}^{+}}^{2}}{\Delta v_{\mbox{\scriptsize C}_{3}\mbox{\scriptsize H}_{2}}^{2}}\right\rangle } & = & 0.90\\
\sqrt{\left\langle \frac{\Delta v_{\mbox{\scriptsize N}_{2}\mbox{\scriptsize H}^{+}}^{2}}{\Delta v_{\mbox{\scriptsize CCS}}^{2}}\right\rangle } & = & 1.05\\
\sqrt{\left\langle \frac{\Delta v_{\mbox{\scriptsize CCS}}^{2}}{\Delta v_{\mbox{\scriptsize C}_{3}\mbox{\scriptsize H}_{2}}^{2}}\right\rangle } & = & 0.97\,.\end{eqnarray*}

It is therefore evident that the differences in mean line width between
ion and neutral species are quite insignificant (compare with equations
(\ref{eq:H13CO+})-(\ref{eq:HCS+})). We should point out, however,
that we have assumed that all the species are coexistent in the clouds.
This is not necessarily true, the authors indeed discuss the lack
of correlation between the column densities of N$_{2}$H$^{+}$ and
those of the neutral species.

It is also worth mentioning that in a recent paper \citet{Crutcher 1999} gives
upper limits for the component of the magnetic field parallel to the
line of sight for a few of the sources studied by \citet{Benson 1998}.
These measurements were obtained using observations of the Zeeman
effect in OH emission lines at 1665 and 1667 MHz with a beam size
of $18\arcsec$ \citep{Crutcher 1993}. The upper limits are all of
the order of $\sim10$ $\mu$G, underlining the relative weakness
of the magnetic field. 

One might then argue that, according to equations (\ref{eq:vz})-(\ref{eq:ndisp}),
perhaps the field strength is too low to cause any narrowing in the
lines profiles of ions. But since we know that this manifestation
should be apparent in the strong magnetic field limit ($\sim0.01$
$\mu$G at a density of $\sim10^{4}$ cm$^{-3}$, which is well below
the typical value measured in the interstellar medium \citep{Heiles 1987}),
it is probably safe to assume that the low level of turbulence and
other flows is the cause for the similarity between the ion and neutral
spectral widths and not the weakness of the field.

\placetable{ta:Benson}

\section{Conclusion.}

We have presented new observational evidence that helps to verify
our assertions that the presence of a sufficiently strong magnetic
field in a weakly ionized and turbulent plasma leads to ions exhibiting
narrower line profiles and significantly suppressed high velocity
wings when compared to neutral lines.

The effect was verified by comparing optically thin transitions of
H$^{13}$CN, H$^{13}$CO$^{+}$ and HCS$^{+}$ made on a sample of
ten high density molecular clouds. The results are in agreement with
our previous observations (Paper I) of probably optically thick lines
detected in OMC1, OMC2, OMC3 and DR21OH.

We also discussed the comparison of the line widths of N$_{2}$H$^{+}$,
CCS, C$_{3}$H$_{2}$ and NH$_{3}$ made on samples of cold and primarily
thermally supported clouds previously published by \citet{Benson 1998}.
We show that the lack of a significant difference between the widths
of ion to neutral lines is exactly what would be expected for this
type of object and therefore confirm a different aspect of our concept
presented in Paper I.

\acknowledgements{We thank Prof. G. A. Blake for directing us to the paper of \citet{Benson 1998}
and Prof. P. M. Solomon for his comments and suggestions. M. Houde's
work was done in part with the assistance of grants from FCAR and
the Département de Physique of the Université de Montréal. The Caltech
Submillimeter Observatory is funded by the NSF through contract AST
9615025. }


\clearpage

\begin{deluxetable}{lccccccccccccccccc}

\tabletypesize{\scriptsize}

\tablecaption{line widths in star formation regions. \label{ta:widths}}

\rotate

\tablecolumns{18}

\tablewidth{0pt}

\tablehead{

\colhead{} & \multicolumn{3}{c}{$\sigma_{v}$ (km/s)\tablenotemark{a}} & 

\colhead{} & \multicolumn{3}{c}{$\sigma_{v}$ (km/s)\tablenotemark{b}} & 

\colhead{} & \multicolumn{4}{c}{$\sigma_{v}$ (km/s)\tablenotemark{c}} & 

\colhead{} & \multicolumn{4}{c}{$\sigma_{v}$ (km/s)\tablenotemark{d}} \\

\cline{2-4} \cline{6-8} \cline{10-13} \cline{15-18} 

\colhead{} & \colhead{HCN} & \colhead{HCO$^{+}$} & 

\colhead{N$_{2}$H$^{+}$} & \colhead{} &

\colhead{HCN} & \colhead{HCO$^{+}$} & 

\colhead{N$_{2}$H$^{+}$} & \colhead{} &

\colhead{H$^{13}$CN} & \colhead{H$^{13}$CO$^{+}$} & 

\colhead{H$_{3}$O$^{+}$} & \colhead{H$_{3}$O$^{+}$} & \colhead{} &

\colhead{H$^{13}$CN} & \colhead{H$^{13}$CO$^{+}$} &  

\colhead{HCS$^{+}$} & \colhead{HCS$^{+}$} \\

\colhead{Source} & \colhead{$ 4-3$} & \colhead{$ 4-3$} & 

\colhead{$ 4-3$} & \colhead{}& \colhead{$ 3-2$} & 

\colhead{$ 3-2$} & \colhead{$ 3-2$} & \colhead{} &

\colhead{$ 4-3$} & \colhead{$ 4-3$} & 

\colhead{$ 3^{+}_{0}-2^{-}_{0}$} & 

\colhead{$ 3^{+}_{2}-2^{-}_{2}$} & \colhead{} &

\colhead{$ 3-2$} & \colhead{$ 3-2$} & 

\colhead{$ 6-5$} & \colhead{$ 5-4$}

}

\startdata

W3 IRS5 & 6.88 & 2.96 & \nodata & & \nodata & \nodata & \nodata & & 

5.25 & 1.51 & 2.38\tablenotemark{f}& 2.38\tablenotemark{f}& & 

4.25 & 1.99 & 1.21 & 1.20 \\

L1551 IRS5 & 1.03 & 0.92 & 0.94 & & \nodata & \nodata & \nodata & & 

\nodata & \nodata & \nodata & \nodata

& & \nodata & \nodata & \nodata & \nodata \\

OMC3-MMS6 & 1.40 & 0.71 & 0.56 & & 1.25 & 0.79 & \nodata & & 

\nodata & 0.40 & \nodata & \nodata

& & 0.98 & 0.47 & 0.51\tablenotemark{e}& 

0.55\tablenotemark{e}\\

OMC1 & \nodata & \nodata & \nodata & & 17.42 & 3.23 & 1.87 & & 

\nodata & \nodata & \nodata & \nodata

& & 8.55 & 1.85 & \nodata & \nodata \\

OMC2-FIR4 & 3.59 & 2.72 & 1.34 & & \nodata & \nodata & \nodata & & 

3.33 & 1.05 & \nodata & \nodata

& & 3.02 & 0.65 & \nodata & \nodata \\

NGC 2071 & 3.98 & 3.69 & 1.18 & & \nodata & \nodata & \nodata & & 

3.12 & 2.01 & \nodata & \nodata

& & \nodata & \nodata & \nodata & \nodata \\ 

NGC 2264 & 1.88 & 1.60 & 1.38 & & \nodata & \nodata & \nodata & & 

1.04 & 0.92 & \nodata & \nodata

& & \nodata & \nodata & \nodata & \nodata \\

M17SWN & 3.29 & 2.96 & 1.82 & & \nodata & \nodata & \nodata & & 

2.82 & 1.91 & \nodata & \nodata

& & 2.19 & 2.03 & \nodata & \nodata \\

M17SWS & 2.04 & 1.84 & 1.32 & & \nodata & \nodata & \nodata & & 

0.89 & 0.69 & \nodata & \nodata

& & 1.51 & 1.19 & \nodata & \nodata \\

DR21OH & 5.75 & 4.61 & 2.08 & & \nodata & \nodata & 2.41 & &

2.95 & 1.93 & \nodata & \nodata

& & 2.82 & 2.04 & 1.86 & 1.72 \\

S140 & 2.69 & 2.14 & 1.34 & & \nodata & \nodata & \nodata & & 

1.50 & 1.36 & \nodata & \nodata

& & 1.42 & 1.12 & 1.09 & 0.90 \\

\enddata

\tablenotetext{a}{Optically thick lines observed with a beam size of $\simeq20\arcsec$.}

\tablenotetext{b}{Optically thick lines observed with a beam size of $\simeq32\arcsec$.}

\tablenotetext{c}{Optically thin lines observed with a beam size of $\simeq20\arcsec$.}

\tablenotetext{d}{Optically thin lines observed with a beam size of $\simeq32\arcsec$.}

\tablenotetext{e}{The corresponding spectra have lower SNR with a line width uncertainty of $\simeq0.1$ km/s.}

\tablenotetext{f}{From \citet{Phillips et al. 1992}. These spectra were obtained with a lower resolution spectrometer, their line width uncertainty is $\simeq0.3$ km/s.}

\end{deluxetable}

\clearpage

\begin{deluxetable}{lrrrcc}

\tabletypesize{\footnotesize}

\tablecaption{Ion to neutral width ratios in star formation regions. \label{ta:ratios}}

\tablecolumns{6}

\tablewidth{0pt}

\tablehead{

\colhead{} & \multicolumn{2}{c}{Coordinates (1950)} & 

\colhead{$v$} & \multicolumn{2}{c}{$\langle$ratio$\rangle$} \\ 

\cline{2-3} \cline{5-6} 

\colhead{Source} & \colhead{RA} & \colhead{DEC} & \colhead{(km/s)} & 

\colhead{thick\tablenotemark{a}} & \colhead{thin\tablenotemark{b}}

}

\startdata

W3 IRS5 & $2^{\mathrm{h}}21^{\mathrm{m}}53\fs3 $ 

& $61\arcdeg52\arcmin21\farcs4$ & $-38.1$ & 0.43 & 0.39 \\

L1551 IRS5 & $4^{\mathrm{h}}28^{\mathrm{m}}40\fs2 $ 

& $18\arcdeg01\arcmin41\farcs0$ & 6.3 & 0.89 & \nodata \\

OMC1 & $5^{\mathrm{h}}32^{\mathrm{m}}47\fs2 $ 

& $-05\arcdeg24\arcmin25\farcs3$ & 9.0 & 0.19 & 0.22 \\

OMC3-MMS6 & $5^{\mathrm{h}}32^{\mathrm{m}}55\fs6 $ 

& $-05\arcdeg03\arcmin25\farcs0$ & 11.3 & 0.51 & 0.48 \\

OMC2-FIR4 & $5^{\mathrm{h}}32^{\mathrm{m}}59\fs0 $ 

& $-05\arcdeg11\arcmin54\farcs0$ & 11.2 & 0.76 & 0.27 \\

NGC 2071 & $5^{\mathrm{h}}44^{\mathrm{m}}30\fs2 $ 

& $00\arcdeg20\arcmin42\farcs0$ & 9.5 & 0.93 & 0.64 \\

NGC 2264 & $6^{\mathrm{h}}38^{\mathrm{m}}25\fs6 $ 

& $09\arcdeg32\arcmin19\farcs0$ & 8.2 & 0.85 & 0.88 \\

M17SWN & $18^{\mathrm{h}}17^{\mathrm{m}}29\fs8 $ 

& $-16\arcdeg12\arcmin55\farcs0$ & 19.6 & 0.90 & 0.81 \\

M17SWS & $18^{\mathrm{h}}17^{\mathrm{m}}31\fs8 $ 

& $-16\arcdeg15\arcmin05\farcs0$ & 19.7 & 0.90 & 0.78 \\

DR21OH & $20^{\mathrm{h}}37^{\mathrm{m}}13\fs0 $ 

& $42\arcdeg12\arcmin00\farcs0$ & $-2.6$ & 0.80 & 0.69 \\

S140 & $22^{\mathrm{h}}17^{\mathrm{m}}40\fs0 $ 

& $63\arcdeg03\arcmin30\farcs0$ & $-7.0$ & 0.80 & 0.85 \\

\enddata

\tablenotetext{a}{From the ratio of HCO$^{+}$ to HCN line width in table \ref{ta:widths}.}

\tablenotetext{b}{From the root mean square of ratios of H$^{13}$CO$^{+}$ to H$^{13}$CN line width in table \ref{ta:widths}.}

\end{deluxetable}

\clearpage

\begin{deluxetable}{lrrcccc}

\tabletypesize{\footnotesize}

\tablecaption{line widths in dense dark clouds. \label{ta:Benson}}


\tablecolumns{7}

\tablewidth{0pt}

\tablehead{

\colhead{} & \multicolumn{2}{c}{Coordinates (1950)} & \colhead{} & \multicolumn{3}{c}{$\Delta v $ (km/s)\tablenotemark{a}} \\

\cline{2-3} \cline{5-7} 

\colhead{Source} & \colhead{RA} & \colhead{DEC} & \colhead{} &

\colhead{N$_{2}$H$^{+}$}& \colhead{CCS}& \colhead{C$_{3}$H$_{2}$} 

}

\startdata

Per 5 & $3^{\mathrm{h}}26^{\mathrm{m}}45\fs5$ & 

$31\arcdeg28\arcmin48\farcs0$ & & 0.37 & 0.27 & 0.43 \\

B5 & $3^{\mathrm{h}}44^{\mathrm{m}}28\fs7$ & 

$32\arcdeg43\arcmin30\farcs0$ & & 0.43 & 0.41 & 0.48 \\

L1498 & $4^{\mathrm{h}}07^{\mathrm{m}}50\fs0$ & 

$25\arcdeg02\arcmin13\farcs0$ & & 0.25 & 0.20 & 0.30 \\

L1495 & $4^{\mathrm{h}}11^{\mathrm{m}}02\fs7$ & 

$28\arcdeg01\arcmin58\farcs0$ & & 0.25 & 0.25 & 0.32 \\

L1527 & $4^{\mathrm{h}}36^{\mathrm{m}}49\fs3$ & 

$25\arcdeg57\arcmin16\farcs0$ & & 0.31 & 0.28 & 0.57 \\

L1512 & $5^{\mathrm{h}}00^{\mathrm{m}}54\fs4$ & 

$32\arcdeg39\arcmin00\farcs0$ & & 0.19 & 0.18 & 0.26 \\ L43E & $16^{\mathrm{h}}31^{\mathrm{m}}46\fs3$ & 

$-15\arcdeg40\arcmin50\farcs0$ & & 0.27 & 0.69 & 0.45 \\

L260 & $16^{\mathrm{h}}44^{\mathrm{m}}22\fs3$ & 

$-09\arcdeg30\arcmin02\farcs0$ & & 0.22 & 0.21 & 0.20 \\

L234E & $16^{\mathrm{h}}45^{\mathrm{m}}22\fs6$ & 

$-10\arcdeg51\arcmin43\farcs0$ & & 0.23 & 0.24 & 0.34 \\

L63 & $16^{\mathrm{h}}47^{\mathrm{m}}21\fs0$ & 

$-18\arcdeg01\arcmin00\farcs0$ & & 0.24 & 0.36 & 0.27 \\

L483 & $18^{\mathrm{h}}14^{\mathrm{m}}50\fs5$ & 

$-04\arcdeg40\arcmin49\farcs0$ & & 0.35 & 0.38 & 0.44 \\

B133 & $19^{\mathrm{h}}03^{\mathrm{m}}25\fs3$ & 

$-06\arcdeg57\arcmin20\farcs0$ & & 0.62 & 0.46 & 0.39 \\

B335 & $19^{\mathrm{h}}34^{\mathrm{m}}33\fs3$ & 

$07\arcdeg27\arcmin00\farcs0$ & & 0.32 & 0.26 & 0.45 \\

L1251E & $22^{\mathrm{h}}38^{\mathrm{m}}10\fs8$ & 

$74\arcdeg55\arcmin50\farcs0$ & & 0.93 &1.03 & 0.85 \\

\enddata

\tablenotetext{a}{FWHM, from \citet{Benson 1998}.}

\end{deluxetable}

\begin{thebibliography}{Zuckerman \& Evans(1974)}
\bibitem[Bachiller(1996)]{Bachiller 1996}Bachiller, R. 1996, \araa, 34, 111
\bibitem[Bachiller(1997)]{Bachiller 1997}Bachiller, R. 1997, in Molecules in astrophysics: probes and processes,
ed. van Dishoeck, E. F. (Dordrecht: Kluwer), 103
\bibitem[Benson et al.(1998)]{Benson 1998}Benson, P. J., Caselli, P., Myers, P. C. 1998, \apj, 506, 743
\bibitem[Choudhuri(1998)]{Choudhuri 1998}Choudhuri, A. R. 1998, The physics of fluids and plasmas, an introduction
for astrophysicists (Cambridge)
\bibitem[Crutcher(1999)]{Crutcher 1999}Crutcher, R. M. 1999, \apj, 520, 706
\bibitem[Crutcher et al.(1993)]{Crutcher 1993}Crutcher, R. M., Troland, T. H., Goodman, A. A., Heiles, C., Kazès,
I., \& Myers, P. C. 1993, \apj, 407, 175
\bibitem[Crutcher et al.(1999)]{Crutcher et al. 1999}Crutcher, R. M., Troland, T. H., Lazareff, B., Paubert, G., Kazès,
I. 1999, \apj, 514, L121
\bibitem[Emerson(1996)]{Emerson 1996}Emerson, D. 1996, Interpreting astronomical spectra (Wiley)
\bibitem[Frisch(1995)]{Frisch 1995}Frisch, U. 1995, Turbulence (Cambridge)
\bibitem[Girart et al.(1999)]{Girart et al. 1999}Girart, J. M., Ho, P. T. P., Rudolph, A. L., Estalella, R., Wilner,
D. J., and Chernin, L. M. 1999, \apj, 522, 921
\bibitem[Heiles(1987)]{Heiles 1987}Heiles, C. 1987, in Interstellar processes, eds. Hollenbach, D. H.,
Thronson Jr, H. A. (Reidel), 171
\bibitem[Houde et al.(2000)]{Houde 2000}Houde, M., Bastien, P., Peng, R., Phillips, T. G., and Yoshida, H.
2000, \apj, in press (Paper I)
\bibitem[Kuiper et al.(1996)]{Kuiper 1996}Kuiper, T. B. H., Langer, W. D., and Velusamy, T. 1996, \apj, 468,
761
\bibitem[Mouschovias(1991a)]{Mouschovias 1991a}Mouschovias, T. Ch. 1991a, in The physics of star formation, eds.
C. J. Lada, N. D. Kylafis (Dordrecht: Kluwer), 61
\bibitem[Mouschovias(1991b)]{Mouschovias 1991b}Mouschovias, T. Ch. 1991b, in The physics of star formation, eds.
C. J. Lada, N. D. Kylafis (Dordrecht: Kluwer), 449
\bibitem[Phillips et al.(1981)]{Phillips 1981}Phillips, T. G., Knapp, G. R., Huggins, P. J., Werner, M. W., Wannier,
P. G., and Neugebauer, G., and Ennis, D. 1981, \apj, 245, 512
\bibitem[Phillips et al.(1992)]{Phillips et al. 1992}Phillips, T. G., van Dishoeck, E. F., and Keene, J. B. 1992, \apj,
399, 533
\bibitem[Shu et al.(1987)]{Shu et al. 1987}Shu, F. H., Adams, F. C., \& Lizano, S. 1987 \araa, 25, 23
\bibitem[Tang et al.(1995)]{Tang et al. 1995}Tang, J. and Saito, S. 1995, \apj, 451, L93
\bibitem[Tennekes \& Lumley(1972)]{Tennekes 1972}Tennekes, H, Lumley, J. L. 1972, A first course in Turbulence, (MIT
Press)
\bibitem[Ungerechts et al.(1997)]{Ungerechts 1997}Ungerechts, H., Bergin, E. A., Goldsmith, P. F., Irvine, W. M., Schloerb,
F. P., \& Snell, R. L. 1997, \apj, 482, 245.
\bibitem[Zuckerman \& Evans(1974)]{Zuckerman 1974}Zuckerman, B., Evans, N. J. II 1974, \apj, 192, L149.
\end{thebibliography}
\end{document}